\documentclass[aps,pre,reprint]{revtex4-1}

\usepackage{bm}
\usepackage{graphicx}
\usepackage{eucal}
\usepackage{mathrsfs}
\usepackage{enumerate}
\usepackage{xcolor}
\usepackage{array}

\newcommand{\comments}[1]{}

\begin{document}

\title{Perturbation expansions at large order: Results for scalar field theories revisited}

\author{Alan J.~McKane}
\affiliation{Theoretical Physics Division, School of Physics and Astronomy,
The University of Manchester, Manchester M13 9PL, United Kingdom}

\begin{abstract}
The question of the asymptotic form of the perturbation expansion in scalar field theories is reconsidered. Renewed interest in the computation of terms in the $\epsilon$-expansion, used to calculate critical exponents, has been frustrated by the differing and incompatible results for the high-order behaviour of the perturbation expansion reported in the literature. We identify the sources of the errors made in earlier papers, correct them, and obtain a consistent set of results. We focus on $\phi^4$ theory, since this has been the most studied and is the most widely used, but we also briefly discuss analogous results for $\phi^N$ theory, with $N>4$. This reexamination of the structure of perturbation expansions raises issues concerning the renormalisation of non-perturbative effects and the nature of the Feynman diagrams at large order, which we discuss.
\end{abstract}
\maketitle

\section{Introduction}
\label{sec:intro}
Forty years ago there was a flurry of activity among field theorists centred on the calculation of the form of the high-order terms in perturbative coupling-constant expansions~\cite{leguillou_1990}. This activity was initiated by the realisation that non-trivial solutions of the field equations---instantons---could be used to systematically calculate the nature of the perturbation expansion at high orders in field theories~\cite{lam_1968,bender_1971,lipatov_1977a}. As is often the case, the activity lasted for only a relatively short period, due in part to technical difficulties that were encountered. For example, extending the approach from scalar field theories to fermionic and gauge theories proved difficult~\cite{parisi_1977a,itzykson_1977b,balian_1978,bogomolny_1978}. A consequence is that there are generations of theorists who know nothing of this work, or if they are aware of it, are unclear as to what results were established.

This state of affairs is by no means unusual in theoretical physics, however in this case these results are becoming more and more relevant to some aspects of current research. This is due to the continuing development of computer-assisted calculation of multiloop Feynman diagrams and in the use of the theory of numbers and single-valued functions on the complex plane, to approach higher-loop calculations (see Refs.~\cite{panzer_2017,schnetz_2018} and references therein). Researchers in these areas are obtaining results which are starting to probe these high-order estimates, and they naturally wish to utilise the results obtained from instanton calculations to extract the best estimates from their perturbative calculations. They have made valiant efforts in this direction (see for example the detailed analysis given in Ref.~\cite{panzer_2017}), but the technical nature of the early papers and the fact that their final results are inconsistent with each other, make it a very difficult task. It was for this reason that they contacted the present author, who was involved in obtaining some of the high-order estimates, in an attempt to see if a consistent set of results could be extracted from these earlier papers. The purpose of the current paper is to do just this.

The technical nature of the papers and the period of time that has passed since they were written, meant that the only realistic way to proceed was to repeat many of the calculations. Fortunately, the papers agree on large parts of the analysis and the differences in final results can be attributed to just two errors---both resulting from the incorrect evaluation of integrals. This may seem rather prosaic, and indeed one of the errors can be described in this way, in that it has no deeper significance. The other error is more subtle, and is connected to the way in which the non-perturbative contributions due to instantons should be renormalised and a conjecture as to which types of diagrams dominate the high-order contributions. These should be matters of general interest to field theorists today, and so a second aim of this paper is to rekindle interest in these questions.

In an attempt to be as clear as possible, and to prevent the paper from becoming too long, we will focus mainly on $\phi^4$ theory, in which most of the work has been done. The errors we refer to only exist in or near $d=4$ dimensions, where the theory is just renormalisable. Calculations carried out directly in $d=3$~\cite{brezin_1978,malatesta_2017} do not have the same difficulties and we will not discuss these calculations here. We will focus our discussion on five papers. The first three were by Lipatov~\cite{lipatov_1977b}, Br\'{e}zin at al.~\cite{BLZ_1977a}, and McKane and Wallace~\cite{mckane_1978}. We will refer to these as the `early papers'. The other two were by McKane et al.~\cite{mckane_1984} and by Komarova and Nalimov~\cite{nalimov_2001}. We will refer to these two papers as the `later papers'. There were other papers, which will be referenced later, but the essential confusion of the subject may be understood with reference to these five papers.

The results from the three early papers are in agreement, however the results of the two later papers are not in agreement with these, nor with each other. Remarkably, these disagreements are not highlighted in the later papers, almost certainly because their origin were not appreciated at the time. The prosaic error mentioned above occurs in the paper of Komarova and Nalimov, and was found through a correspondence between the authors of Ref.~\cite{panzer_2017} and this paper in the last year or so~\cite{nalimov_2017,panzer_2016}. Here we will simply point out the origin of the error, and indicate how it can be corrected. With this correction, the paper of Komarova and Nalimov~\cite{nalimov_2001} agrees with that of McKane et al.~\cite{mckane_1984}. Therefore we are left with the three early papers giving one set of results and the two later papers a different set of results.

The discrepancy between the two sets of papers has its origins in the way that the non-perturbative part of the vertex functions, generated by the instanton, is renormalised. It is fortunate that there is no disagreement about the calculation of this non-perturbative contribution, so we do not have to enter into a discussion of the instanton calculation itself, only the renormalisation of the result of this calculation.

The outline of the paper is as follows. In Sec.~\ref{sec:background_disagreements} an introduction to the general method of obtaining high-order estimates from instanton contributions is given. The source of the disagreement between the early papers and the later ones is explained, and the case made that the later treatments are the correct ones. In Sec.~\ref{sec:HOB_RG_functions} the high-order behaviour of renormalisation group (RG) functions is obtained, as are those of the fixed point and critical exponents in $4-\epsilon$ dimensions. This calculation was previously preformed in Ref.~\cite{mckane_1984}, but the authors of Ref.~\cite{nalimov_2001} were doubtful of its correctness due to the `non-perturbative' nature of the analysis. We show that a more mundane treatment gives the same result. In Sec.~\ref{sec:dominant_graphs} we give an overview of two topics which are related to our analysis, and which we believe deserve further investigation. One relates to the nature of the dominant diagrams at high order and the other is the possible effect renormalons~\cite{lautrup_1977,tHooft_1977b} may have on the analysis which we have presented. We conclude in Sec.~\ref{sec:conclude}. There are two appendices: one on the analogous calculations for $\phi^N$ theory when $N > 4$ and the other on the derivation of various mathematical results used in the text.

\section{Background and Conflicting Results}
\label{sec:background_disagreements}
This section is divided into two parts. In the first, the basic ideas behind the calculation of high-order estimates in perturbation theory are outlined. The purpose is to serve as a quick review of the method for those who are not familiar with it, but also to establish notation and standard results which will be used later in the paper. The second part identifies the errors made in previous papers.

\subsection{Methodology}
\label{sec:methods}
The calculation falls into two distinct parts.

\subsubsection{Calculation of the imaginary part of vertex functions}
\label{sec:imag_part_calc}
The field theory being investigated is studied for values of the coupling constant where it is unstable. In the case where the interaction is $g\phi^4/4$, this means taking the coupling constant, $g$, to be negative. The instability of the theory at negative coupling means that the vertex functions develop an imaginary part; this is what we wish to calculate. It can be found by using what is in effect the method of steepest descent, where the extremum in function space is the instanton, which will be denoted by $\phi_{\rm c}(x)$. 

The details of the calculation are given in the papers already mentioned (we will follow Ref.~\cite{mckane_1978}), and as we have already stressed, there is no need to reproduce them here. However there is one point which needs to be explained, since it lies at the heart of the inconsistencies between the papers. Instantons typically have a position, and so expanding about one of them breaks the translational invariance of the theory. However in $d=4$ dimensions the theory has a larger symmetry, the instanton also has a scale as well as a position, and so an extra zero mode is created through the breaking of this symmetry. The collective coordinate introduced to deal with this~\cite{zittartz_1966,langer_1967,gervais_1975} will be denoted by $\lambda$ (representing dilatations). It is this extra integral over $\lambda$, not present in field theories below their upper critical dimension, that is the root of the errors in earlier papers.

After this short summary we will now simply state the result found for the imaginary part of the four point vertex function in $\phi^4$ theory, which is generated when $g < 0$ by the instability of the theory and which is found by an instanton calculation:
\begin{eqnarray}
& & \left. \mathrm{Im} \Gamma^{(4)}_{\rm b}\left( q_i \right)\right|_{{\rm arg}g=\pi} = - C_{\rm b} \int^\infty_0 \frac{d\lambda}{\lambda} \lambda^\epsilon \left( - \frac{\lambda^\epsilon A}{g} \right)^{(5+d)/2} \nonumber \\
& & \times \ \exp{ \left( \frac{\lambda^\epsilon A}{g} \right) } \prod^4_{i=1} \left[ \left( \frac{q^2_i}{\lambda^2} \right) \tilde{\phi}_{\rm c}\left( \frac{q_i}{\lambda}\right)\right] \left[ 1 + \mathcal{O}\left( g,\epsilon\right) \right],
\label{starting_point}
\end{eqnarray}
where
\begin{eqnarray}
\label{A}
A &=& \frac{8}{3} \pi^2 + \mathcal{O}\left( \epsilon \right), \\
\label{C_b}
C_{\rm b} &=& 2^{-1/2} \pi^{-3} \exp{ \left( \frac{3}{\epsilon} + \frac{3\zeta'(2)}{\pi^2} - \frac{7}{2}\gamma - \frac{15}{4} \right) }, \\
\tilde{\phi}_{\rm c}\left(q\right) &=& 2^{d/2}\pi^{(d-2)/2} 3^{1/2}\left| q \right|^{(2-d)/2}K_{(d-2)/2}\left( q \right).
\label{phi_tilde}
\end{eqnarray}
These are Eqs.~(37)-(39) of Ref.~\cite{mckane_1978} and Eqs.~(2.1)-(2.4) of Ref.~\cite{mckane_1984}. Equivalent results are to be found in Refs.~\cite{lipatov_1977b} and \cite{BLZ_1977a}, but these papers used a momentum cut-off to regularise the divergences, rather than dimensional regularisation.

A number of points need to be made relating to Eqs.~(\ref{starting_point})-(\ref{phi_tilde}). First, some definitions. The function $K$ is a modified Bessel function, $d=4-\epsilon$, $\gamma$ is Euler's constant and $\zeta'(2)$ is the derivative of the zeta function evaluated at the integer $2$. The result is expressed in momentum space, and $q_i$, $i=1,\ldots,4$ are the four momenta associated with this four point vertex function. We have specified that arg $g=\pi$. If arg $g=-\pi$, the expression is minus the right-hand side of Eq.~(\ref{starting_point}); there is a cut along the negative $g$ axis with a discontinuity equal to twice this expression. Finally, the subscript `b' on $C$ denotes it is a bare quantity: note the $3/\epsilon$ in the exponential. Similarly the subscript `b' on $\Gamma^{(4)}$ denotes it is a bare vertex function.

Equation (\ref{starting_point}) is the result of calculating the effects of fluctuations about the instanton. The quantity in the exponential is the `classical' result and is essentially the action of the instanton. The prefactor is obtained from evaluating a determinant, which comes from Gaussian fluctuations about the instanton. The order $g$ corrections come from higher-order fluctuations about the instanton.

We end this section with a few explanatory and technical points. The first of these relates to the constant $A$ defined in Eq.~(\ref{A}). As discussed in Ref.~\cite{mckane_1984}, the order $\epsilon$ correction to $A$ does not contribute to the results for the high-order behaviour of the critical exponents at the order at which we are working. Therefore the undesirable expansion of $A$ as a power series in $\epsilon$ within the exponential, which would give rise to $\mathcal{O}(\epsilon/g)$ terms, may be avoided, and so in the following the value of $A$ is taken to be its value when $\epsilon = 0$. Secondly, we wish to mention two other papers on the topic of this paper, but which have no direct impact on the central message of the current work. The first is the instanton calculation of Drummond and Shore~\cite{drummond_1979} which, in part, involved a discussion on the correct form of Eq.~(\ref{starting_point}). Their result agreed with that given by this equation, and so for our current purposes there is no need to pursue this further. The other paper is a next-to-leading order calculation of the instanton contribution~\cite{nalimov_2005}. We have not carried out such a calculation and therefore do not comment on it further. Finally, we will make an attempt to pinpoint disagreements with previous papers, but since Lipatov~\cite{lipatov_1977b} and Br\'{e}zin at al.~\cite{BLZ_1977a} agree, and both used a momentum cut-off to regulate the theory, we will make comparisons to specific results in the latter paper only, but our comments of course also apply to the former.

\subsubsection{Dispersion relation}
\label{sec:disp_rel}
Once the imaginary part of the vertex function has been found for $g<0$ a dispersion relation allows estimates of the high orders in perturbation theory to be found. Specifically the existence of a branch cut for $g < 0$ leads to
\begin{eqnarray}
\Gamma^{(4)}_{\rm b}\left( q_i, g \right) &=& \frac{1}{\pi}\int^0_{-\infty} \frac{dg'}{g' - g } \left. \mathrm{Im} \Gamma^{(4)}_{\rm b}\left( q_i, g' \right)\right|_{{\rm arg}g'=\pi} \nonumber \\
&\sim& \sum_K g^K \frac{1}{\pi}\int^0_{-\infty} \frac{dg'}{\left( g' \right)^{K+1}} \left. \mathrm{Im} \Gamma^{(4)}_{\rm b}\left( q_i, g' \right)\right|_{{\rm arg}g'=\pi}. \nonumber \\
\label{dis_rel}
\end{eqnarray}
While rigorous results are available to justify this approach when $d < 4$~\cite{eckmann_1975,feldman_1976,magnen_1977}, the analytic properties of the vertex functions in the complex coupling constant plane when $d=4$ are less clear. This will be discussed further in Sec.~\ref{sec:bubble_graphs}.

The main thrust of the current paper is that various previous papers disagree on the result for $\mathrm{Im} \Gamma^{(4)}_{\rm b}( q_i )|_{{\rm arg}g=\pi}$ \textit{after} the $\lambda$-integral has been carried out. However all agree that as a function of $g$ it has the general form 
\begin{equation}
\left. \mathrm{Im} \Gamma^{(4)}_{\rm b}\left( q_i \right)\right|_{{\rm arg}g=\pi} = \mathcal{A}_1\left( - g \right)^{-\xi}\exp{ \left( \frac{\mathcal{A}_0}{g} \right) } \left[ 1 + \mathcal{O}\left( g \right) \right],
\label{gen_form_Gamma_4}
\end{equation}
where $\xi$ is related to the number of symmetries that the choice of instanton breaks and $\mathcal{A}_0$ and $\mathcal{A}_1$ are constants. Substitution of Eq.~(\ref{gen_form_Gamma_4}) into the right-hand side of Eq.~(\ref{dis_rel}), and performing the integral gives for the coefficient of $g^K$ large $K$:
\begin{eqnarray}
& & \left( - 1 \right)^{K+1}\frac{\mathcal{A}_1}{\pi} \left( \frac{1}{\mathcal{A}_0} \right)^{(K+\xi)} \Gamma\left( K + \xi \right)\left[ 1 + \mathcal{O}\left( K^{-1}\right) \right] = \nonumber \\
& & \left( - 1 \right)^{K+1} K! K^{\xi - 1} \left( \frac{1}{\mathcal{A}_0} \right)^K \frac{\mathcal{A}_1}{\pi \mathcal{A}_0^\xi} \left[ 1 + \mathcal{O}\left( K^{-1}\right) \right].
\label{generic_HO}
\end{eqnarray}
Previous authors~\cite{lipatov_1977b,BLZ_1977a,mckane_1978,mckane_1984,nalimov_2001} agree on the structure shown in Eq.~(\ref{generic_HO}), and on the value of $\mathcal{A}_0$ and $\xi$ (found from a purely `classical' calculation). Where they differ is on the value of $\mathcal{A}_1$. This may seem to be just an overall constant, but the early papers obtain a finite result and the later papers find it has a pole in $\epsilon$. The problem lies in (i) performing the $\lambda$ integral, so as to move from Eq.~(\ref{starting_point}) to Eq.~(\ref{gen_form_Gamma_4}), and (ii) renormalising the theory. We will now discuss these two points, and then identify the source of the disagreements.

\subsection{Identification of previous errors}
\label{sec:errors}
As indicated the errors originate from the incorrect evaluation of two integrals. We begin by discussing the error caused through the evaluation of the $\lambda$ integral in Eq.~(\ref{starting_point}). 

\subsubsection{Renormalisation and the $\lambda$ intergral}
\label{sec:error1}
Although all three of the earlier papers agree, the intermediate steps differ, since two of them~\cite{lipatov_1977b,BLZ_1977a} used a cut-off to regularise the divergences and the other~\cite{mckane_1978} used dimensional regularisation. We will use dimensional regularisation here, since we find it clearer, it is more familiar to readers and it allows for the use of minimal subtraction (MS). However our comments below also relate to the calculation with a cut-off. In this case a regulator of the form $(1/2\Lambda^2)\phi(\partial^2)^2\phi$ could be added to the action, which would result in the $\lambda$-integral in Eq.~(\ref{starting_point}) being replaced by~\cite{BLZ_1977a}
\begin{equation}
\int^\infty_0 \frac{d\lambda}{\lambda} e^{-\rho^2 \lambda^2/\Lambda^2}\,\prod^4_{i=1} \left[ \left( \frac{q^2_i}{\lambda^2} \right) \tilde{\phi}_{\rm c}\left( \frac{q_i}{\lambda}\right)\right],
\label{BLZ_lambda_int}
\end{equation}
where $\rho^2= 1/2 \int (\partial^2)^2 \phi^2_{\rm c}(x)d^4 x$.

Although the discrepancies can be resolved by a careful analysis of the $\lambda$ integral, they also involve wider questions which relate to how non-perturbative contributions such as these should be renormalised. So we begin with a discussion which touches on these points and give general arguments as to why we would expect the $\lambda$ integral to diverge in $d=4$ dimensions.

The three early papers found that a one-loop renormalisation of the coupling constant was sufficient to render the $\lambda$ integral in Eq.~(\ref{starting_point}) (or equivalently in Eq.~(\ref{BLZ_lambda_int})) finite. This was rationalised as follows. The instanton is a smooth extended object, and so one would not expect that the ultra-violet divergences found by expanding about it would be any different to those found by expanding about $\phi=0$. On the other hand, one can argue that the overall constant $\mathcal{A}_1$ which appears in Eq.~(\ref{gen_form_Gamma_4}), and which contains the result of carrying out the $\lambda$ integral, also appears in the coefficient of $g^K$ for large $K$ (see Eq.~(\ref{generic_HO})). In this setting we would expect it to contain divergences which could not be eliminated through a one-loop renormalisation. To make this last statement clearer we can imagine working with a $\phi^4$ theory with an interaction 
$g_{abcd}\phi_a\phi_b\phi_c\phi_d$, where $g_{abcd}$ is a fourth rank tensor of some symmetry group. Typically at each order there will be some novel contractions of $g_{abcd}$ not found at lower orders, and divergences involving these will not be able to be cancelled by those found in one-loop diagrams.

A second argument which suggests that the $\lambda$ integral diverges involves consideration of $\phi^6$ theory in $d=3$. This has many of the features of $\phi^4$ theory in $d=4$, the crucial one is that they are both just renormalisable. The difference is that the Gaussian fluctuations about the instanton in $\phi^6$ theory give a finite result in $d=3$, and so do not need to be renormalised. In Appendix \ref{sec:appA} we summarise the basic features of the instanton calculation for $\phi^N$ theory with $N>4$, and give the (finite) result for the Gaussian fluctuations about the instanton in the $N=6$ case. From the argument of the last paragraph we might still expect a divergence in the $\lambda$ integral, and indeed the authors who reported a finite $\lambda$ integral for $\phi^4$, do report a divergent integral in $\phi^6$ theory (with a cut-off---see Eq.~(73) of Ref.~\cite{BLZ_1977a}, and in dimensional regularisation---see Eq.~(6.30) of Ref.~\cite{mckane_PhD}). It seems bizarre to expect the diagrams that contribute at $K^{\rm th}$ order in $\phi^6$ theory to diverge, but not those in $\phi^4$ theory, and this difference to be a consequence of a coupling-constant renormalisation which was carried at to lowest order in $\phi^4$, but not in $\phi^6$.

 After these general arguments, we now show the existence of a divergence in the $\lambda$ integral in Eq.~(\ref{starting_point}) by direct evaluation. This is correctly carried out in Ref.~\cite{mckane_1984}, but we will give it again here, with a little more detail, given it is central to the disagreements in the literature. First we note that the integral converges for small $\lambda$ --- even in the limit $\epsilon \to 0$ --- due to the exponential decay of the modified Bessel function $K$ for large arguments. However, we will see that that the integral diverges logarithmically for large $\lambda$ as $\epsilon \to 0$. 

To explore this, we use the small argument asymptotic form for the Bessel function (see Eq.~(\ref{K_effective}) in Appendix \ref{sec:app_B1}) in Eq.~(\ref{phi_tilde}), which implies that 
\begin{equation}
\frac{q^2_i}{\lambda^2}\tilde{\phi}\left( \frac{q_i}{\lambda} \right) \sim 2^2 3^{1/2} \pi \left( \frac{q_i}{\lambda} \right)^{\epsilon} \left[ 1 + \mathcal{O}\left( \epsilon \right) \right].
\label{instanton_large_lambda}
\end{equation}
Splitting up the $\lambda$ integration as $\int^\mu_0 + \int^\infty_\mu$, where $\mu$ is an arbitrary momentum scale, then the divergence, if it exists, is contained in 
\begin{eqnarray}
& & - C_{\rm b} \left( 2^2 3^{1/2} \pi \right)^4 \prod^4_{i=1} \left( q_i \right)^{\epsilon}\,\int^\infty_\mu \frac{d\lambda}{\lambda} \lambda^{-3\epsilon} \left( - \frac{\lambda^\epsilon A}{g} \right)^{(5+d)/2} \nonumber \\
& & \times \ \exp{ \left( \frac{\lambda^\epsilon A}{g} \right) } \left[ 1 + \mathcal{O}\left( g,\epsilon\right) \right].
\label{contained_in}
\end{eqnarray}
The integral may be evaluated using the following result
\begin{eqnarray}
& & \int^\infty_\mu \frac{d\lambda}{\lambda} \lambda^{a\epsilon} \lambda^{\epsilon}\,\exp{ \left[ - \left( \frac{\lambda^\epsilon A}{|g|} \right) \right] } \nonumber \\
& & = \frac{1}{\epsilon} \frac{|g|}{A} \mu^{a\epsilon}\,\exp{ \left[ - \left( \frac{\mu^\epsilon A}{|g|} \right) \right] } \left[ 1 + \mathcal{O}\left( g \right) \right],
\label{integral_form}
\end{eqnarray}
for some positive constant $a$ (this result can be proved by a change of variable $x=\lambda^\epsilon A/|g|$ and then an integration by parts). 

Using Eq.~(\ref{integral_form}), the expression (\ref{contained_in}) becomes
\begin{eqnarray}
& & - \frac{2^8 3^2 \pi^4 C_{\rm b}}{\epsilon} \left( - \frac{g}{A} \right)\left( - \frac{A\mu^\epsilon}{g} \right)^{(5+d)/2} \nonumber \\
& & \times \ \exp{ \left( \frac{\mu^\epsilon A}{g} \right) }\prod^4_{i=1} \left( \frac{q_i}{\mu} \right)^\epsilon \left[ 1 + \mathcal{O}\left( g, \epsilon \right) \right].
\label{result_intermediate}
\end{eqnarray}
The factors $(q_i/\mu)$ can be expanded in $\epsilon$ and give typical finite contributions $\ln (q_i/\mu)$. We are then left with the pole term
\begin{eqnarray}
\left. \mathrm{Im} \Gamma^{(4)}_{\rm b}\left( q_i \right)\right|_{{\rm arg}g=\pi} &=& \frac{2^8 3^2 \pi^4 C_{\rm b}}{\epsilon} \left( \frac{g}{A} \right)\left( - \frac{A\mu^\epsilon}{g} \right)^{(5+d)/2} \nonumber \\
& & \times \ \exp{ \left( \frac{\mu^\epsilon A}{g} \right) }\left[ 1 + \mathcal{O}\left( g, \epsilon \right) \right].
\label{result_canonical}
\end{eqnarray}
This is now in the canonical form shown in Eq.~(\ref{gen_form_Gamma_4}) and we can begin renormalisation. However we stress again that the integral was formally divergent in $d=4$ and that this divergence is regulated by $\epsilon=4-d$, and manifests itself by a pole in $\epsilon$ for small $\epsilon$. The pole term is all that will interest us here, since we are using the MS scheme.

Why then did the early papers report a different result? To understand this point it is sufficient to focus on the expression
\begin{equation}
C \exp{ \left( \frac{3}{\epsilon} \right) }\,\int^\infty_\mu \frac{d\lambda}{\lambda} \lambda^{-3\epsilon} \left( - \frac{\lambda^\epsilon A}{g} \right)^{(5+d)/2}\,\exp{ \left( \frac{\lambda^\epsilon A}{g} \right) },
\label{toy_integral}
\end{equation}
which we denote by $I_\epsilon(g)$. This is the quantity in Eq.~(\ref{contained_in}), with finite factors incorporated in the constant $C$, but with the $3/\epsilon$ in the exponential of $C_{\rm b}$ (see Eq.~(\ref{C_b})) made explicit.

Proceeding as above we may evaluate the integral using Eq.~(\ref{integral_form}) to find
\begin{equation}
I_\epsilon(g) = \frac{C}{\epsilon}\left( - \frac{g}{A} \right)\left( - \frac{A\mu^\epsilon}{g} \right)^{(5+d)/2}\exp{ \left( A \left[ \frac{\mu^\epsilon}{g} + \frac{3}{A\epsilon} \right] \right) },
\label{lambda_int_first_way}
\end{equation}
up to corrections of order $g$. Now the one loop renormalisation (found by a conventional one-loop Feynman graph calculation) is:
\begin{equation}
\frac{\mu^\epsilon}{g} = \frac{1}{g_{\rm r}} - \frac{3}{A\epsilon},
\label{renorm_dim_reg}
\end{equation}
where the renormalisation of the coupling constant has been written in this way, so it is obvious that the $3/\epsilon$ in the exponential is cancelled. The lower case subscript of $g$ means `one-loop renormalised'. Using Eq.~(\ref{renorm_dim_reg}) gives the following expression for Eq.~(\ref{lambda_int_first_way}):
\begin{equation}
\frac{C \mu^\epsilon}{\epsilon}\left( - \frac{g_{\rm r}}{A} \right)\left( - \frac{A}{g_{\rm r}} \right)^{(5+d)/2}\exp{ \left( \frac{A}{g_{\rm r}} \right) }\left[ 1 + \mathcal{O}\left( g_{\rm r}, \epsilon \right) \right].
\label{renorm_first_way}
\end{equation}
Therefore a one-loop renormalisation eliminates the pole in the exponential, but a pole in front of the whole expression remains.

Now we describe the procedure which was in effect carried out in the early papers. The renormalisation was carried out first, to obtain for the expression in Eq.~(\ref{toy_integral}) 
\begin{eqnarray}
& & C \exp{ \left( \frac{3}{\epsilon} \right) }\,\int^\infty_\mu \frac{d\lambda}{\lambda} \lambda^{-3\epsilon} \left( - \frac{\lambda^\epsilon A}{\mu^\epsilon g_{\rm r}} \right)^{(5+d)/2} \nonumber \\
& & \times \ \exp{ \left\{ \left( \frac{\lambda}{\mu} \right)^\epsilon A \left[ \frac{1}{g_{\rm r}} - \frac{3}{A\epsilon} \right] \right\} }\left[ 1 + \mathcal{O}\left( g_{\rm r}, \epsilon \right) \right] \nonumber \\
& & = C \int^\infty_\mu \frac{d\lambda}{\lambda} \lambda^{-3\epsilon} \left[ - \left( \frac{\lambda}{\mu} \right)^\epsilon \frac{A}{g_{\rm r}} \right]^{(5+d)/2}\,\exp{ \left\{ \left( \frac{\lambda}{\mu} \right)^\epsilon \frac{A}{g_{\rm r}} \right\} } 
\nonumber \\
& & \times \ \exp{ \left( \frac{3}{\epsilon} \left[ 1 - \left( \frac{\lambda}{\mu} \right)^\epsilon \right] \right) }\left[ 1 + \mathcal{O}\left( g_{\rm r}, \epsilon \right) \right].
\label{renorm_second_way}
\end{eqnarray}
The key step is when the argument of the second exponential in Eq.~(\ref{renorm_second_way}) is expanded out:
\begin{equation}
\frac{3}{\epsilon} \left[ 1 - \left( \frac{\lambda}{\mu} \right)^\epsilon \right] \ \longrightarrow -3\ln \left( \frac{\lambda}{\mu} \right) + \mathcal{O}\left( \epsilon \right).
\label{erroneous}
\end{equation}
If this is used in the integral in Eq.~(\ref{renorm_second_way}) it produces a factor of $(\lambda/\mu)^{-3}$, which causes the integral to converge in the limit $\epsilon \to 0$. So setting $\epsilon = 0$ one finds a finite result for $\mathcal{A}_1$. 

The integral in Eq.~(\ref{renorm_second_way}), which is expressed in terms of $g_{\rm r}$, is still of the form given in Eq.~(\ref{integral_form}), and so when evaluated still diverges like $\epsilon^{-1}$ as $\epsilon \to 0$. It is only the replacement given in Eq.~(\ref{erroneous}) that results in an integral which has a finite $\epsilon \to 0$ limit. This is the erroneous step made in the early papers. 

In Ref.~\cite{mckane_1978} the convergence factor $\lambda^{-3}$ is not immediately in evidence, since the coupling constant is renormalised at the scale $\lambda$, however substitution of Eq.~(44) into Eq.~(42) of that paper brings it out. There is also an explicit comment that the asymptotic freedom of the theory (in $d=4$ and for $g<0$) provides this convergent factor. Although the introduction of the coupling constant renormalised at the scale $\lambda$, $g_{\rm r}(\lambda)$, seems natural, it takes us further away from being able to carry out the $\lambda$ integral. To evaluate the integral we have to relate $g_{\rm r}(\lambda)$ to $g_{\rm r}$ renormalised at a fixed scale $\mu$, $g_{\rm r}(\mu)$, but this generates the factor on the left-hand side of Eq.~(\ref{erroneous}), and the subsequent erroneous step of replacing it by the right-hand side of Eq.~(\ref{erroneous}).

In Ref.~\cite{BLZ_1977a} the first relevant step is in the unnumbered equation after Eq.~(85) (the terms in this equation appear in the argument of an exponential, where they are multiplied by $-3/\epsilon$). This split is exactly as was carried out in Eq.~(\ref{renorm_second_way}), the only difference being that the renormalisation in Ref.~\cite{BLZ_1977a} was carried out by the introduction of counterterms in the action. The step in Eq.~(\ref{erroneous}) gives rise to the factor in Eq.~(87) of Ref.~\cite{BLZ_1977a} (with $n=1$) which gives the finite integral in Eq.~(88) of that paper. In this part of the calculation $\epsilon=4-d$ has been used to regulate the divergences, and the momentum cut-off removed. Since the integral diverges, a consistent treatment would be to let $\epsilon \to 0$ with the cutoff in place, and to presumably obtain a $\ln \Lambda$ divergence. However since we will work within dimensional regularisation there is no need to follow this route.

Although we will not pursue the calculations described in the early papers further, since we believe they are incorrect, we will make a few comments which may make the results given in these papers a little easier to understand. Firstly, since the result of the $\lambda$ integral was finite, the restriction of the range of integration in Eq.~(\ref{contained_in}) to $\lambda > \mu$ no longer holds, and the whole range of $\lambda$ has to be included. Therefore in the four-point vertex function, the Bessel function $K_1$ appears to the fourth power (see Eq.~(\ref{phi_tilde})) in the final result for the overall factor $\mathcal{A}_1$. In the correct approach only the pole in $\epsilon$ is of interest, and the Bessel functions only enter through their asymptotic forms given in Eq.~(\ref{instanton_large_lambda}). It was the existence of `Bessel moments', such as $\int^\infty_0 y^6 [ K_1(y) ]^4 dy$, in the early papers, but not in the later ones, that caused so much confusion to recent researchers.  

As we will discuss in Sec.~\ref{sec:HOB_RG_functions}, the divergence which remains in the four-point vertex function, after a one-loop renormalisation has been carried out, is removed by a coupling constant renormalisation at $K^{\rm th}$ order. The second point we wish to mention is that such a coupling constant renormalisation was carried out in the early papers, even though performing the $\lambda$ integral did not result in a divergence. This was then a finite renormalisation, and was implemeted through renormalisation conditions, for example defining a renormalised coupling constant at the value of the four-point vertex at a symmetry point, as shown in Eq.~(46) of Ref.~\cite{mckane_1978} and Eq.~(75) of Ref.~\cite{BLZ_1977a}. This results in a subtraction of the $\lambda$-integral at the symmetry point as shown in Eq.~(88) of Ref.~\cite{BLZ_1977a}. As remarked before, a divergence was found in $\phi^6$ theory in $d=3$ and so the renormalisation given in Eq.~(76) of Ref.~\cite{BLZ_1977a} does remove an infinity, unlike in $\phi^4$ theory.

Another point that is worth mentioning is that because the $\lambda$ integral involved complicated terms, such as the fourth power of the Bessel function $K_1$, its evaluation was delayed. However since finding the $\beta$-function involved finding the derivative of the $\lambda$ integral with respect to the momentum scale $\mu$, an integration by parts allowed the $\lambda$ integral to factor out, as shown in Eq.~(47) and (48) of Ref.~\cite{mckane_1978}, leaving the `Bessel moment' displayed in Eq.~(53) of that paper.

Finally, we stress again that the early papers show that the calculations carried out directly in $d=4$---with a cut-off~\cite{lipatov_1977b,BLZ_1977a}---and those carried out in $d=4-\epsilon$~\cite{mckane_1978,drummond_1979}, agree, after a coupling constant renormalisation and after the regularisation has been removed (the cut-off taken to infinity or $\epsilon$ taken to zero). The correspondence between the two calculations was established at the time~\cite{mckane_1978,drummond_1979}. However, as we have discussed, the error they made originated in the evaluation of the $\lambda$ integral, and this occurred if the calculation was carried out directly in $d=4$ or if dimensional regularisation was used: in both cases a spurious factor of $\lambda^{-3}$ was generated which rendered the $\lambda$ integral finite.

We believe that the discussion in this section resolves the discrepancies in the evaluation of the $\lambda$ integral in the various papers in the literature. We will discuss the correct treatment in more detail in Sec.~\ref{sec:HOB_RG_functions}, but now we go on to discuss the second integral in which an error appeared.

\subsubsection{The constant $U_{\rm reg}$}
\label{sec:error2}
This second error is far easier to discuss, since it has no additional implications. In Eq.~(13) of Ref.~\cite{nalimov_2001} a constant $U$ is defined, the `regular' (i.e. non-divergent) part of which is 
\begin{eqnarray}
U_{\rm reg} &\equiv& \frac{1}{16\pi^2}\,\int d^{d}k\,v(k) \left\{ \left[ \psi(1) + \ln(4\pi) + 2 \right] \right. \nonumber \\
&-& \left. 2\ln |k| \right\} v(-k),
\label{U_reg_def}
\end{eqnarray}
where $\psi(1)$ is the digamma function with argument $1$ and $v(k)$ is essentially the Fourier transform of the square of the instanton solution. The error occurs in the part of $U_{\rm reg}$ which involves $\ln |k|$ and was identified~\cite{nalimov_2017,panzer_2016} in correspondence between the authors of Refs.~\cite{nalimov_2001} and \cite{panzer_2017}. Here we will simply indicate where the error occurred (in Appendix \ref{sec:app_B2}), and give the corrected result. 

On page 344 of Ref.~\cite{nalimov_2001} the value of $U_{\rm reg}$ in $d=4$ and with $n=1$ is given as 
\begin{equation}
\frac{3}{2}\left[ - \gamma - \ln \pi + 2 \right],
\label{wrong_U_reg}
\end{equation}
whereas a correct evaluation~\cite{nalimov_2017,panzer_2016}---see Appendix \ref{sec:app_B2}---gives
\begin{equation}
\frac{3}{2}\left[ \gamma + \ln \pi + \frac{5}{3} \right].
\label{correct_U_reg}
\end{equation}
The difference $3[ \gamma + \ln \pi - 1/6]$ is just the factor required~\cite{nalimov_2017,panzer_2016} to give agreement with Ref.~\cite{mckane_1984}.

\section{Asymptotic forms for RG functions}
\label{sec:HOB_RG_functions}
In Sec.~\ref{sec:errors} we pinpointed the errors which previously existed in the literature which, when corrected, give consistent results for $\mathrm{Im} \Gamma^{(4)}_{\rm b}( q_i )|_{{\rm arg}g=\pi}$ when expressed in the form given by Eq.~(\ref{generic_HO}). Here we begin with this result and outline the derivation of the high-order estimates for the RG $\beta$ function and of the critical exponents in $4-\epsilon$ dimensions. The correct form in the case of the critical exponents has already been given in Ref.~\cite{mckane_1984}, but there are several reasons for discussing the calculation here.

First, the reason why the authors of Ref.~\cite{nalimov_2001} did not question the correctness of their calculation when it was found to differ from that given in Ref.~\cite{mckane_1984} was that they believed the latter to be incorrect, due to the use of `non-perturbation renormalisation' in that paper. This phrase refers to the delayed use of a dispersion relation; instead of starting from Eq.~(\ref{generic_HO}), the calculation continues in the $g<0$ regime, finding exponentially small imaginary parts for the renormalisation constants, the RG functions, the non-trivial fixed point (for $\epsilon < 0$), and finally for the critical exponents. A dispersion relation is finally used in the complex $\epsilon$ plane to find the high-order behaviour of the critical exponents. 

Here we will carry out the calculation in a more conventional manner, starting with Eq.~(\ref{generic_HO}) and working in the physical regime $g>0$. Therefore there are no imaginary parts for the renormalisation constants, RG functions etc. We obtain identical results as those found using a `non-perturbation renormalisation', albeit less elegantly, as we shall see. This vindicates the use of the methods of Ref.~\cite{mckane_1984}.

Another reason for revisiting the calculation is that the method of Ref.~\cite{mckane_1984} gave high-order estimates for the critical exponents, but not for the RG functions. They could be deduced from the information given after some calculation, but the asymptotics of the RG $\beta$ function is of interest, and we explicitly give the estimates in this section.

We begin the calculation from Eq.~(\ref{result_canonical}), but written in terms of the one-loop renormalised coupling constant $g_{\rm r}$ defined in Eq.~(\ref{renorm_dim_reg}):
\begin{eqnarray}
\left. \mathrm{Im} \Gamma^{(4)}_{\rm r}\left( q_i \right)\right|_{{\rm arg}g=\pi} &=& - \frac{2^8 3^2 \pi^4 C_{\rm r}}{\epsilon}\mu^\epsilon \left( - \frac{A}{g_r} \right)^{(3+d)/2} \nonumber \\
&\times& \exp{ \left( \frac{A}{g_r} \right) }\left[ 1 + \mathcal{O}\left( g_r, \epsilon \right) \right],
\label{result_one_loop}
\end{eqnarray}
where
\begin{eqnarray}
C_{\rm r} &=& C_{\rm b}\exp{ \left( - \frac{3}{\epsilon} \right) } \nonumber \\
&=& 2^{-1/2} \pi^{-3} \exp{ \left( \frac{3\zeta'(2)}{\pi^2} - \frac{7}{2}\gamma - \frac{15}{4} \right) }.
\label{C_r}
\end{eqnarray}
This was the form schematically written down in Eq.~(\ref{renorm_first_way}). It has the general structure shown in Eq.~(\ref{gen_form_Gamma_4}) with $\xi = (3+d)/2$, and so we can find the high-order behaviour by directly using Eq.~(\ref{generic_HO}). In this way we find that the coefficient of $g_r^K$ in $\Gamma^{(4)}_{\rm r}\left( q_i \right)$ for $K$ large is given by
\begin{equation}
\frac{2^8 3^2 \pi^3 C_{\rm r}}{\epsilon} \mu^\epsilon\,\left( - 1 \right)^K\,K!\,K^{(d+1)/2}\,\left( \frac{1}{A} \right)^K \left[ 1 + \mathcal{O}\left(\frac{1}{K} \right) \right].
\label{coeff_of_g_tothe_K_one_loop}
\end{equation}

In the above we appear to have been disregarding the wavefunction renormalisation. This is because there is no wavefunction renormalisation at one loop in $\phi^4$ theory. However, it is also the case that the wavefunction renormalisation does not come in at $K^{\rm th}$ order to leading order in $K$. To see this we note that the imaginary part of the wavefunction renormalisation constant is down by a power of $g$ on the four-point vertex (see, for instance, Eq.~(2.24) of Ref.~\cite{mckane_1984}), which means that the coefficient of $g_{\rm r}^K$ in the perturbative expansion is down by a factor of $K^{-1}$ on the perturbative expansion of the four-point vertex. So when we multiply $\Gamma^{(4)}_{\rm r}\left( q_i \right)$ by the wavefunction renormalisation factor, only $\mathcal{O}(K^{-1})$ factors potentially change in the coefficient of $g_{\rm r}^K$. 

We have used the notation $g_{\rm r}$ for the renormalisation of the bare coupling constant at one loop; we shall write $g_{\rm R}$ for the fully renormalised coupling constant (i.e., at $K$ loops for $K$ large). Using Eq.~(\ref{coeff_of_g_tothe_K_one_loop}) we therefore have, using MS,
\begin{eqnarray}
& & g_{\rm R}(\mu) = g_{\rm r}(\mu) + \ldots + \ldots - g^K_{\rm r}(\mu)\,\frac{2^7 3 \pi^3 C_{\rm r}}{\epsilon} \nonumber \\
&\times& \left( - 1 \right)^K K!\,K^{(d+1)/2}\,\left( \frac{1}{A} \right)^K \left[ 1 + \mathcal{O}\left(\frac{1}{K} \right) \right] + \ldots.
\label{full_renorm_pert}
\end{eqnarray}
In terms of the bare coupling constant $g$ this reads:
\begin{eqnarray}
& & g_{\rm R}\left( \mu \right) = g \mu^{-\epsilon} - \frac{9}{8\pi^2 \epsilon} g^2 \mu^{-2\epsilon} + \ldots - g^K \mu^{-K\epsilon}\,\frac{2^7 3 \pi^3 C_{\rm r}}{\epsilon}
\nonumber \\
& & \times \left( - 1 \right)^K\,K!\,K^{(d+1)/2}\,\left( \frac{1}{A} \right)^K \left[ 1 + \mathcal{O}\left(\frac{1}{K} \right) \right] + \ldots.
\label{g_R_pert}
\end{eqnarray}
To obtain Eq.~(\ref{g_R_pert}) we have to note that the only contribution to the $\mathcal{O}(g^K)$ term comes from the $\mathcal{O}(g_{\rm r}^K)$, since all other terms are down by factors of $K$, because of the $K!$ growth.

\subsection{High order behaviour of the RG $\beta$ function}
\label{sec:HOB_beta_function}
We can now differentiate Eq.~(\ref{g_R_pert}) with respect to $\mu$ at fixed $g$ to obtain the $\beta$ function,
\begin{eqnarray}
\beta(g_{\rm R}) &=& \left. \mu\left( \frac{\partial }{\partial \mu} \right)\right|_{g}  g_{\rm R}(\mu) \nonumber \\
&=& -\epsilon g \mu^{-\epsilon} + \frac{9}{4\pi^2} g^2 \mu^{-2\epsilon} + \ldots + g^K \mu^{-K\epsilon}\,2^7 3 \pi^3 C_{\rm r} \nonumber \\
&\times& \left( - 1 \right)^K\,K!\,K^{(d+3)/2}\,\left( \frac{1}{A} \right)^K \left[ 1 + \mathcal{O}\left(\frac{1}{K} \right) \right] + \ldots. \nonumber \\
\label{beta_full_1_pert}
\end{eqnarray}
This may be written as :
\begin{equation}
-\epsilon g_{\rm R} + \frac{9}{8\pi^2} g^2 \mu^{-2\epsilon} + \ldots,
\label{beta_full_2_pert}
\end{equation}
the $\mathcal{O}(g^K)$ part coming from the $g_{\rm R}$ in the first term again being a factor of $K$ down on the $\mathcal{O}(g^K)$ term already present (due to the $K$ brought down after the $\mu$ differentiation to get the $\beta$ function). We therefore arrive at
\begin{eqnarray}
\beta(g_{\rm R}) &=& -\epsilon g_{\rm R} + \frac{9}{8\pi^2} g_{\rm R}^2 + \ldots + \ldots + g_{\rm R}^K\,2^7 3 \pi^3 C_{\rm r} \nonumber \\
&\times& \left( - 1 \right)^K\,K!\,K^{(d+3)/2}\,\left( \frac{1}{A} \right)^K \left[ 1 + \mathcal{O}\left(\frac{1}{K} \right) \right] + \ldots. \nonumber \\
\label{beta_final_pert}
\end{eqnarray}

So, in summary, if we write
\begin{equation}
\beta(g_{\rm R}) = -\epsilon g_{\rm R} + \sum^\infty_{K=2} \beta_K g_{\rm R}^K,
\label{beta_funct_pert}
\end{equation}
then, substituting for $A$ from Eq.~(\ref{A}),
\begin{equation}
\beta_K = C_\beta \left( - 1 \right)^K K!\,K^{7/2}\,\left( \frac{3}{8\pi^2} \right)^K \left[ 1 + \mathcal{O}\left(\frac{1}{K} \right) \right],
\label{HOB_beta_1}
\end{equation}
for large $K$, where 
\begin{equation}
C_\beta = 2^{13/2}3 \exp{ \left( \frac{3\zeta'(2)}{\pi^2} - \frac{7}{2}\gamma - \frac{15}{4} \right) }.
\label{C_beta}
\end{equation}
We have also set $d=4$, in order to obtain the result in MS. Equation (\ref{HOB_beta_1}) should be contrasted with Eq.~(96) of Ref.~\cite{BLZ_1977a}, and Eq.~(53) of Ref.~\cite{mckane_1978}, all of which contain the Bessel moment $\int^\infty_0 y^6 [ K_1(y) ]^4 dy$, coming from the fact that the entire range of the $\lambda$ integral has contributed to $C_\beta$, rather than just the large $\lambda$ divergence.

\subsection{High order behaviour of the fixed point $g^*_{\rm R}$}
\label{sec:HOB_fixed_point}
We denote the non-trivial fixed point of the RG $\beta$ function in $d=4-\epsilon$ dimensions by $g^*_{\rm R}$. At low orders it is given in MS by~\cite{mckane_1984}
\begin{equation}
g^*_{\rm R} = \frac{8}{9}\pi^2 \epsilon\left[ 1 + s_1\epsilon + \mathcal{O}\left( \epsilon^2 \right) \right], \ \ \ s_1 = \frac{17}{27},
\label{g_R_two_loops}
\end{equation}
where the result has been given explicitly to two loops. 

We begin from Eq.~(\ref{beta_final_pert}), setting $\beta(g^*_{\rm R})=0$:
\begin{eqnarray}
0 &=& -\epsilon + \frac{9}{8\pi^2} g^*_{\rm R} + \ldots - g_{\rm R}^{*\,K}\,2^7 3 \pi^3 C_{\rm r} \left( \frac{3}{8\pi^2} \right) \left( - 1 \right)^K \nonumber \\
&\times& K!\,K^{9/2}\,\left( \frac{3}{8\pi^2} \right)^K \left[ 1 + \mathcal{O}\left(\frac{1}{K} \right) \right] + \ldots,
\label{fp_from_pert}
\end{eqnarray}
where a factor of $g^*_{\rm R}$ has been cancelled throughout, and so where the $\mathcal{O}\left( g_{\rm R}^{*\,K} \right)$ term is actually the $\mathcal{O}\left( g_{\rm R}^{*\,(K+1)} \right)$ term with a factor of $g^*_{\rm R}$ cancelled out.

We would expect to substitute the perturbative expression for $g^*_{\rm R}$ given in Eq.~(\ref{g_R_two_loops}) into the factor $g_{\rm R}^{*\,K}$ in Eq.~(\ref{fp_from_pert}) and then identify the high-order behaviour of $g^*_{\rm R}$ using the early terms in Eq.~(\ref{fp_from_pert}). However it is not as simple as this. To see this we note from Eq.~(\ref{g_R_two_loops}), that in raising $g^*_{\rm R}$ to the power $K$, we produce the term $[ 1 + s_1\epsilon + \mathcal{O}( \epsilon^2 ) ]^K$, which gives terms such as $K\epsilon$ when expanded out. This is the same order as the naive leading term found from $g_{\rm R}^{*\,(K+1)}$. Obviously this term gives an $\epsilon^{(K+1)}$ factor and so is not of immediate interest to us, but it also suggests that the $g_{\rm R}^{*\,(K-1)}$ will give a contribution to $\epsilon^K$ through such a mechanism. 

To investigate this further, we raise the expression for $g_{\rm R}^*$ given in Eq.~(\ref{g_R_two_loops}) to the power $(K-\ell)$. Here $\ell=0,1,2,\ldots$ and we will assume that $\ell$ is such that $(K-\ell)$ can still be thought of as large. One finds
\begin{eqnarray}
& & \left[ \frac{8}{9}\pi^2 \epsilon \right]^{(K-\ell)}\,\left[ 1 + s_1\epsilon + \mathcal{O}\left( \epsilon^2 \right) \right]^{(K-l)} =
\nonumber \\
& & \left[ \frac{8}{9}\pi^2 \epsilon \right]^{(K-\ell)}\,\exp{(K-\ell)\ln\left[ 1 + s_1\epsilon + \mathcal{O}\left( \epsilon^2 \right) \right] }.
\nonumber \\
\label{K_minus_ell_contrib}
\end{eqnarray}
Expanding out the logarithm, we see that the only term liable to give relevant $\epsilon^K$ terms comes from $\exp{(K-\ell)s_1\epsilon}$. Expanding this exponential out to get the relevant $\epsilon^K$ term gives the contribution
\begin{eqnarray}
& & \left[ \frac{8}{9}\pi^2 \epsilon \right]^{(K-\ell)}\,\frac{\left[ (K-\ell)s_1\epsilon \right]^{\ell}}{\ell !} \sim \left[ \frac{8}{9}\pi^2 \epsilon \right]^{(K-\ell)}\,\frac{\left[ K s_1\epsilon \right]^{\ell}}{\ell !} \nonumber \\
& & \sim \left[ \frac{8}{9}\pi^2 \epsilon \right]^K\,K^\ell\,\frac{\left[ 9 s_1/8\pi^2 \right]^{\ell}}{\ell !}.
\label{K_minus_ell_only_contrib}
\end{eqnarray}
Putting in the factors that multiply $g_{\rm R}^{*\,(K-\ell)}$ in Eq.~(\ref{fp_from_pert}) and inserting the value of $s_1$ gives, after some algebra,
\begin{equation}
- 2^4 3^2 \pi C_{\rm r}\,K!\,K^{9/2}\,\left( - \frac{1}{3} \right)^K\,\frac{\left[ - 17/9 \right]^{\ell}}{\ell !}\,\epsilon^K.
\label{relevant_fp_from_pert}
\end{equation}
Summing up the terms from $\ell=0,1,\ldots$ gives a factor of $e^{-(17/9)}$, and balancing Eq.~(\ref{relevant_fp_from_pert}) against the early term $9/8\pi^2 g^*_R$ from Eq.~(\ref{fp_from_pert}) gives for the coefficient of $\epsilon^K$ in the fixed point $g^*_{\rm R}$ to be
\begin{equation}
2^7 \pi^3 C_{\rm r}\,K!\,K^{9/2}\,\left( - \frac{1}{3} \right)^K\,e^{-(17/9)}\,\epsilon^K.
\label{Kth_order_fp}
\end{equation}
The whole procedure sketched out above is cumbersome and inelegant, as compared to the technique of non-perturbative renormalisation in Ref.~\cite{mckane_1984}, where the equivalent calculation (starting with Eq.~(3.10)) takes only a few lines, and is far better controlled. 

If we write
\begin{equation}
g^*_{\rm R} = \sum^\infty_{K=1} \left( g^*_{\rm R} \right)_{K} \epsilon^K,
\label{g_star_sum}
\end{equation}
then using Eq.~(\ref{C_r}) one finds, for $K$ large, that
\begin{equation}
\left( g^*_{\rm R} \right)_{K} = \bar{C}_{\rm g} K!\,\left[ -1/3 \right]^K K^{9/2}\left[ 1 + \mathcal{O}\left( \frac{1}{K} \right) \right], 
\label{g_star_pert}
\end{equation}
where
\begin{equation}
\bar{C}_{\rm g} = 2^{13/2} \exp{ \left( \frac{3\zeta'(2)}{\pi^2} - \frac{7}{2}\gamma - \frac{203}{36} \right) }.
\label{C_g_pert}
\end{equation}
The constant has been denoted by $\bar{C}_{\rm g}$, rather than simply $C_{\rm g}$, to accord with the notation of Ref.~\cite{mckane_1984} where the bar signified that these were found in coefficients of $\epsilon^K$, rather than in the high-order estimates before the dispersion relation had been used.

\subsection{High order behaviour of critical exponents in $d=4-\epsilon$}
\label{sec:HOB_critical_exponents}
The critical exponents of the theory are found from the anomalous dimensions of $\phi$ and $\phi^2$ which are defined by
\begin{eqnarray}
\gamma_{\phi}(g_{\rm R}) &=& \left. \mu\left( \frac{\partial }{\partial \mu} \right)\right|_{g}  \ln Z_{\phi} \nonumber \\
\gamma_{\phi^2}(g_{\rm R}) &=& - \left. \mu\left( \frac{\partial }{\partial \mu} \right)\right|_{g}  \ln Z_{\phi^2},
\label{anom_dim}
\end{eqnarray}
where $Z_\phi$ and $Z_{\phi^2}$ are the renormalisation constants found from the renormalisation of the field and the $\phi^2$ operator respectively~\cite{amit_1978}. Substitution of the fixed point value of $g_{\rm R}$ into these functions gives the critical exponents $\eta$ and $\nu^{-1}$~\cite{amit_1978}
\begin{equation}
\eta = \gamma_{\phi}(g^*_{\rm R})\ ; \ \nu^{-1} - 2 = - \gamma_{\phi^2}(g^*_{\rm R}),
\label{ident_crit_expo}
\end{equation}
as power series in $\epsilon$. 

It might be thought that substituting the expression for $g^*_{\rm R}$ into the functions $\gamma_{\phi}(g_{\rm R})$ and $\gamma_{\phi^2}(g_{\rm R})$ that there might possibly be contributions from the high-order terms in $g^*_{\rm R}$ (given in Eq.~(\ref{g_star_pert})) and by the low-order terms (given in Eq.~(\ref{g_R_two_loops})) substituted into the high-order terms in the functions themselves. However it turns out that the latter contributions are absent; as discussed in both the early and later papers (see, for instance, Ref.~\cite{mckane_1984}), the renormalisation constants $Z_\phi$ and $Z_{\phi^2}$, and so the functions $\gamma_{\phi}(g_{\rm R})$ and $\gamma_{\phi^2}(g_{\rm R})$, have contributions which do not contribute to leading order. Therefore to find the high-order behaviour of the critical exponents one only needs to know the low order perturbation results for $\gamma_{\phi}(g_{\rm R})$ and $\gamma_{\phi^2}(g_{\rm R})$.

There is no one-loop contribution to $\gamma_{\phi}(g_{\rm R})$, so the dominant contribution is from the two loop result. One has that~\cite{mckane_1984}
\begin{eqnarray}
\gamma\left( g_{\rm R} \right) &=& \frac{3}{2(8\pi^2)^2} g^2_{\rm R} + \mathcal{O}\left( g^3_{\rm R} \right), \nonumber \\
\gamma_{\phi^2}\left( g_{\rm R} \right) &=& \frac{3}{8\pi^2} g_{\rm R} + \mathcal{O}\left( g^2_{\rm R} \right).
\label{anom_dim_low_order}
\end{eqnarray}
Therefore the $K$th order term ($K$ large) in the $\epsilon$ expansion for $\gamma_{\phi^2}(g^*_{\rm R})$ is from Eq.~(\ref{g_star_pert}) and Eq.~(\ref{anom_dim_low_order})
\begin{equation}
2^{-3} 3 \pi^{-2} \bar{C}_{\rm g} K!\,\left[ -1/3 \right]^K K^{9/2}\left[ 1 + \mathcal{O}\left( \frac{1}{K} \right) \right]. 
\label{gamma_phi_squared_HOB}
\end{equation}
In the case of $\gamma_{\phi}(g_{\rm R})$, which begins at order $g^2_{\rm R}$, the dominant contribution will come from a cross-term between the $\mathcal{O}(\epsilon)$ term and the order $\mathcal{O}(\epsilon^{(K-1)})$ term in $g^{*\,2}_{\rm R}$, giving for the coefficient of $\epsilon^K$ in $\gamma(g^*_{\rm R})$:
\begin{eqnarray}
& & \frac{3}{2^7 \pi^4} \frac{16\pi^2}{9} \bar{C}_{\rm g} K!\,\left[ -1/3 \right]^{(K-1)} K^{7/2}\left[ 1 + \mathcal{O}\left( \frac{1}{K} \right) \right] 
\nonumber \\
&=& - 2^{-3} \pi^{-2}\,\bar{C}_{\rm g} K!\,\left[ -1/3 \right]^K K^{7/2}\left[ 1 + \mathcal{O}\left( \frac{1}{K} \right) \right].
\label{gamma__HOB}
\end{eqnarray}
From the identification of the critical exponents in Eq.~(\ref{ident_crit_expo}), we see that the coefficients of $\epsilon^K$ in $\eta(\epsilon)$ and $\nu^{-1}(\epsilon)$ for $K$ large are
\begin{equation}
\eta_K = - \bar{C}_{\eta} K!\,\left[ -1/3 \right]^K K^{7/2}\left[ 1 + \mathcal{O}\left( \frac{1}{K} \right) \right],
\label{HOB_eta}
\end{equation}
and
\begin{equation}
\nu^{-1}_K = - \bar{C}_{\nu^{-1}} K!\,\left[ -1/3 \right]^K K^{9/2}\left[ 1 + \mathcal{O}\left( \frac{1}{K} \right) \right],
\label{HOB_nu}
\end{equation}
where
\begin{equation}
\bar{C}_{\nu^{-1}} = 3 \bar{C}_{\eta} = \frac{2 ^{7/2} 3}{\pi^2}\,\exp{ \left( \frac{3\zeta'(2)}{\pi^2} - \frac{7}{2}\gamma - \frac{203}{36} \right) }.
\label{bar_relations}
\end{equation}
This agrees with the results given in Ref.~\cite{mckane_1984} (Eqs.~(4.6)-(4.10)), although there they are also given for the $O(n)$ theory, and in addition the correction to scaling exponent $\omega = \beta'(g^*_{\rm R})$ is given. Our aim here was simply to show that these results can be obtained within perturbation theory and for physical values of the coupling constant. On a more technical note, we mention that the corrections in Ref.~\cite{mckane_1984} were given to be of order $\ln K/K$, rather than of order $1/K$ as given here. This is a consequence of setting $d=4$ in the definition of the $\beta$ function (\ref{C_beta}), rather than keeping the $d$-dependence as in Ref.~\cite{mckane_1984}. The choice we have made here seems to us to be more consistent with the MS scheme which we are using.

\section{Which graphs dominate at high orders?}
\label{sec:dominant_graphs}
In previous sections we have described specific calculations, identified disagreements and errors in previous calculations and rectified them. This short section has a different character; here we briefly discuss various conjectures that have been made, or proofs given, relating to the type of graphs which dominate the high order behaviour of the vertex functions. The discussion naturally falls into two parts. First, a consideration of which graphs give the pole in $\epsilon$ found in Eq.~(\ref{coeff_of_g_tothe_K_one_loop}), and second, whether the existence of graphs which may not be captured by the steepest descent calculation, spoil the predictions given in Sec.~\ref{sec:HOB_RG_functions}. We will conclude that both of these questions are still open, and so we do not reach any definitive conclusions. However, we feel that at least a brief summary of the status of these questions is required when writing on the nature of the pertubation expansion in field theories.

\subsection{Totally irreducible (primitive) graphs}
\label{sec:totally_irr_graphs}
There are statements in the literature which state that the pole in $\epsilon$ found in Eq.~(\ref{coeff_of_g_tothe_K_one_loop}) can be attributed to the totally irreducible, that is, the primitive graphs, at $K^{\rm th}$ order. For example, in the text after Eq.~(73) of Ref.~\cite{BLZ_1977a}: ``Finally, the leading diagrams at order $K$ give a single power of $\ln \Lambda$, they are those which do not involve any divergent subgraph; i.e., they are the completely irreducible diagrams''. This was written in relation to $\phi^{2N}$ theory in $d=2N/(N-1)$ dimensions for $N>2$, so for example, $\phi^6$ theory in $d=3$ dimensions. Another example is on page 1865 of Ref.~\cite{mckane_1984}: ``In the context of high-order estimates in the perturbative series, we interpret the extra pole in $\epsilon$ as the one produced by the totally irreducible diagrams at high orders. These diagrams are known to be the dominant ones at $K^{\rm th}$ order for $K$ large for $d < 4$ and moreover diverge only like $1/\epsilon$''. This relates to $\phi^4$ theory. 

Since there are no citations given to substantiate these claims, their status is uncertain. The second of the quotes given above was written by the present author, but he recalls only that there was a general belief in the correctness of the assertion at the time. Certainly one can see why the absence of $1/\epsilon^{\ell}$, $\ell > 1$ divergences might lead one to formulate such a conjecture. Added to this are the experiences of those who have calculated critical exponents with the $\epsilon$-expansion; it has been a recurring theme of those performing these calculations that the totally irreducible diagrams give a disproportionally large contribution to the critical exponents within the $\epsilon$-expansion~\cite{BLZN_1973,wallace_1976,dittes_1978}. In the recent six- and seven-loop calculations it was estimated that the primitive diagrams contribute $69\%$ of the minimally subtracted $\beta$ function at six loops and $78\%$ at seven loops~\cite{panzer_2017}. The actual fraction of graphs which are primitive (in the four point vertex function) is quite small ($10/627  = 0.0159...$  at six loops and $44/3794 = 0.0116...$  at seven loops~\cite{panzer_2017}), and approaches $\exp{ (-9/2) } =0.0111... $ in the limit when the loop number goes to infinity. This is just the ratio of (unlabelled) isomorphism classes of primitive graphs to tadpole free one-particle irreducible graphs, and if one includes symmetry factors, the ratio becomes $\exp{ (-9/4) }$~\cite{borinsky_2017a,borinsky_2017b,panzer_2016}. So it is their contribution that is large, and the idea of them dominating at large order may well be correct.

It is also interesting to note that since the contributions of primitive diagrams to the $\beta$ function are scheme independent, the conjecture that they dominate would seem to imply that the MS result (\ref{HOB_beta_1}) gives the asymptotic behaviour in any renormalisation scheme.

\subsection{Renormalons}
\label{sec:bubble_graphs}
While the nature of the dominant graphs found in an instanton/steepest descent calculation, discussed in Sec.~\ref{sec:totally_irr_graphs}, is interesting, and may be useful in obtaining estimates at large orders, another source of dominant graphs may be more important. These are ``bubble'' graphs which lead to renormalon singularities, found very soon after the early papers on high-order behaviour came out~\cite{lautrup_1977, tHooft_1977b, parisi_1977c,parisi_1978, parisi_1979, bergere_1984}. These potentially give extra singularities on the \textit{positive} $g$-axis, and could invalidate the use of the dispersion relation described in Sec.~\ref{sec:disp_rel}.

These singularities exist only for $d=4$; for $d<4$ the use of a dispersion relation as discussed in Sec.~\ref{sec:disp_rel} is valid. The situation in $d=4$ is still unclear. It is known that the Borel radius is at least equal to the renormalon expected position~\cite{david_1988}, but it appears that little progress has been made on this question in recent years. The consensus seems to be that renormalon singularities probably do exist in vertex functions in $d=4$, although there are arguments and rigorous work which suggest that they are absent, or are not manifest, in some tensor and matrix field theories~\cite{rivasseau_2015,grosse_2018}. In addition, the renormalon singluarities may be absent from the RG functions, if we make use of the following argument~\cite{mckane_1984}. In the MS scheme the RG functions in $d=4-\epsilon$ dimensions have the form $\beta(g_{\rm R}) = -\epsilon g_{\rm R} + \beta^{\{ 4\} }(g_{\rm R})$, $\gamma_\phi(g_{\rm R}) = \gamma^{\{ 4\} }_\phi(g_{\rm R})$ and $\gamma_{\phi^2}(g_{\rm R}) = \gamma^{\{ 4\} }_{\phi^2}(g_{\rm R})$, where the superscripts indicate that these are the RG functions with $d=4$. Therefore if the RG functions are free of singularities in $4 - \epsilon$ dimensions, they are also free of singulaties in $d=4$, \textit{as long as MS is used}. If this is correct, then so are the results given in Sec.~\ref{sec:HOB_RG_functions}, since all calculations were made in $4-\epsilon$ dimensions using the MS scheme. 

Clearly more work needs to be carried out on the topics discussed in this section. Here our aim has not been to present new results, but merely to summarise the situation as we see it, in the hope of stimulating renewed interest in these questions.

\section{Conclusion}
\label{sec:conclude}
The main purpose of this paper has been to identify the errors in several of the key papers which obtained estimates of the high-order terms in perturbative expansions in quantum field theories, to correct them, and to present a consistent set of results, which can be utilised in current research. We concentrated on single component $\phi^4$ theory, since our aim was to present the sources of error in as clear a way as possible. The extension to $n$-component fields with an $O(n)$ symmetry can be found in Ref.~\cite{mckane_1984}, where the results for the $n$-component case are correctly given. The analogous calculation for $\phi^N$ theory with $N>4$ was discussed in Appendix \ref{sec:appA}, although the central error of the $\phi^4$ calculation cannot occur in this case. This is due to the fact that the Gaussian fluctuations about the instanton do not produce a divergence, and so do not need to be renormalised, and hence there is no possibilty of generating a spurious factor consisting of a power of $\lambda$, which can be used to make the dilatation integral convergent.

This leaves $\phi^3$ theory, which is well defined for imaginary coupling constant or in the multicomponent case in certain limits, both of which have applications~\cite{mckane_1981}. A four-loop calculation has recently been carried out~\cite{gracey_2015}, and reliable estimates of high-order behaviour would be useful in this case too. There are two currently available calculations in the literature. The first~\cite{mckane_1979} utilises the method we have shown to be erroneous, but it is straightforward to obtain a corrected result by carrying out the dilatation integral using the result in Eq.~(\ref{integral_form}), and then proceeding as in Sec.~\ref{sec:HOB_RG_functions}. The second~\cite{nalimov_2014} does not have the erroneous convergence factor in the integration over the dilatation parameter $\lambda$, however the authors of this paper were apparently unaware of the earlier paper Ref.~\cite{mckane_1979}, and so do not compare their results to it. It would be useful to reconcile these two approaches, but we do not pursue this here, since it presents no new features, and would add to the length of an already long paper.

One of the conclusions of the paper is that the results presented in Ref.~\cite{mckane_1984} are correct. These were criticised~\cite{nalimov_2001} for using ``non-perturbative renormalisation'', and it was assumed that this was the probable reason for the disagreement with the results presented in Ref.~\cite{nalimov_2001}. The phrase ``non-perturbative renormalisation'' refers to methodology whereby the entire vertex function for arg$\,g = \pi$ (that is, the complex function consisting of its real and imaginary parts) is renormalised. This generates renormalisation constants which are complex, which in turn leads to renormalisation group functions, such as the $\beta$-function, which are complex. This procedure seems very reasonable to us, as it did when Ref.~\cite{mckane_1984} was written, and provides an elegant way of obtaining the high-order behaviour of the critical exponents within the $\epsilon$-expansion. As far as we are aware, no work has been carried out on the rigorous justification of this procedure, but in Sec.~\ref{sec:HOB_RG_functions} the calculation was carried out in a more pedestrian fashion, which gave the same results.

Other open questions, relating to the types of diagram which dominate at large order, are discussed in Sec.~\ref{sec:dominant_graphs}. We believe that some of these issues are capable of being resolved, and hope that, in addition to providing a consistent set of results, this paper will stimulate further research into these questions.

\begin{acknowledgments}
I wish to thank David Broadhurst and Erik Panzer for their initial questions, and later comments, on the inconsistent results in the literature, which led to my gradual reimmersion into the subject. They also encouraged me to write this paper, and together with John Gracey provided numerous useful comments on a draft version. I also wish to thank David Wallace for discussion of our original papers and Carl Bender, Giorgio Parisi and Vincent Rivasseau for useful correspondence.
\end{acknowledgments}

\appendix

\section{The calculation for $\phi^N$ theory, with $N > 4$}
\label{sec:appA}
Although this paper is mainly concerned with $\phi^4$ theory, $\phi^N$ ($N > 4$) is also of interest, especially $\phi^6$ since, as explained in Sec.~\ref{sec:error1}, earlier papers are in agreement that a divergence occurs when evaluating the dilational ($\lambda$) integral. In this appendix we give a very brief outline of the calculation which leads to the analogue of Eq.~(\ref{starting_point}) and give the key results from the renormalisation process. The result of evaluating the determinant of Gaussian fluctuations is given explicitly for $\phi^6$ theory.

The interaction term will be taken to be $g\phi^{N}/N$ and the theory will be studied in $d=d_c - \epsilon$ dimensions where $d_c = 2N/(N-2)$ is the critical dimension where the theory is just renormalisable (the subscript c used here, should not be confused with the subscript c on the $\phi$ field which denotes that it is a `classical' contribution). The case where $N$ is an even integer was investigated by Lipatov~\cite{lipatov_1977b} and Br\'{e}zin et al~\cite{BLZ_1977a}, where potential divergences were regulated using a cut-off. Subsequently dimensional regularisation was used and both odd $N$ and even $N$ studied~\cite{mckane_PhD}. We shall follow the procedure of Ref.~\cite{mckane_PhD} here; note that Br\'{e}zin et al~\cite{BLZ_1977a} denoted the interaction as $\phi^{2N}$, rather than $\phi^N$, since they only investigated theories where the interaction was an even power of $\phi$. 

We need to study the vertex function $\Gamma^{(M)}$ only in the case $M=N$; this vertex function is logarithmically divergent in $d=d_c$ dimensions. If $M>N$, the vertex functions are convergent, if $2 < M < N$ any integrals are zero in dimensional regularisation and if $M=2$ there is a wavefunction renormalisation but, as for $\phi^4$ it does not contribute to the high-order behaviour at the order which we are working. Performing an instanton calculation along exactly the same lines as for $\phi^4$ theory~\cite{mckane_1978} gives for the case of even $N$~\cite{mckane_PhD}
\begin{eqnarray}
& & \left. \mathrm{Im} \Gamma^{(N)}_{\rm b}\left( q_i \right)\right|_{{\rm arg}g=\pi} = - C_{\rm b}(N) \int^\infty_0 \frac{d\lambda}{\lambda} \lambda^{(N-2)\epsilon/2} \nonumber \\
&\times& \left[ \lambda^\epsilon A(N) \left( - \frac{1}{g} \right)^{2/(N-2)} \right]^{(d+1+N)/2} \nonumber \\
&\times& \exp{ \left[ - \lambda^\epsilon A(N) \left( - \frac{1}{g} \right)^{2/(N-2)} \right] } \nonumber \\
&\times& \prod^N_{i=1} \left[ \left( \frac{q^2_i}{\lambda^2} \right) \tilde{\phi}^{(N)}_c\left( \frac{q_i}{\lambda}\right)\right] \left[ 1 + \mathcal{O}\left( g^{2/(N-2)},\epsilon\right) \right] ,
\label{starting_point_phi_N_even}
\end{eqnarray}
where
\begin{eqnarray}
\label{A(N)}
A(N) &=& \left[ \frac{8N}{\left( N - 2 \right)^2} \right]^{2/(N-2)}\frac{4\pi^{d_c/2}}{\left( N - 2 \right)} \frac{\Gamma(d_c/2)}{\Gamma(d_c)} + \mathcal{O}\left( \epsilon \right), \nonumber \\ \\
\label{C_b(N)}
C_{\rm b}(N) &=& \left[ \frac{2}{\pi\left( 3N - 2 \right)} \right]^{(d_c + 1)/2} C_2(N),\\
\tilde{\phi}^{(N)}_{\rm c}\left(q\right) &=& \left( - 1 \right)^N B(N)\left( 2\pi \right)^{-\epsilon/2}\left| q \right|^{-1 + (\epsilon/2)}K_{1 - (\epsilon/2)}\left( q \right). \nonumber \\ 
\label{phi_tilde(N)}
\end{eqnarray}
In Eq.~(\ref{phi_tilde(N)}),
\begin{equation}
B(N) \equiv \sqrt{\frac{8\Gamma(d_c)\pi^{d_c/2}}{\Gamma^2(d_c/2)\Gamma((d_c - 2)/2)}}.
\label{B(N)}
\end{equation}
One can check that setting $N=4$ in Eqs.~(\ref{starting_point_phi_N_even})-(\ref{B(N)}) gives Eqs.~(\ref{starting_point})-(\ref{phi_tilde}) of the main text, apart from the factor $C_2(N)$, which comes from the evaluation of the Gaussian fluctuations about the instanton, and which has been explicitly evaluated in the $N=4$ case. It will be given below for $N=6$. All this is for $N$ even; for $N$ odd an identical result holds but multiplied by an overall factor of $-1/2$ and with the imaginary part evaluated when ${\rm arg}g=0$ rather than ${\rm arg}g=\pi$. 

The procedure now parallels that given in the main text for $\phi^4$ theory. The $\lambda$ integral may be carried out using Eq.~(\ref{integral_form}) and gives a simple pole in $\epsilon$ for all $N > 4$, just as it did for $N=4$. Introducing the dimensionless bare coupling constant $\bar{g} = g \mu^{-(N-2)\epsilon/2}$ and applying the dispersion relation Eq.~(\ref{dis_rel}) when $N$ is even, gives the pole term of the coefficient of $\bar{g}^K$ in $\Gamma^{(N)}(q_i,\bar{g})$ to be
\begin{eqnarray}
&&\left( - 1 \right)^K C_{\rm b}(N)\frac{(N-2)\mu^{(N-2)\epsilon/2}}{2\pi\epsilon}\,\left[ A(N) \right]^{-(N-2)K/2} \left[ B(N) \right]^N \nonumber \\
&& \times \ \Gamma\left\{ \left[ K(N - 2) + (d-1+N) \right]/2 \right\}\,\left[ 1 + \mathcal{O}\left( \epsilon, K^{-1} \right) \right]. \nonumber \\
\label{dis_rel_N_even_result}
\end{eqnarray}

The pole in $\epsilon$ is clearly visible in Eq.~(\ref{dis_rel_N_even_result}). A similar analysis can be carried out when $N$ is odd, although in this case a slightly different dispersion relation has to be used~\cite{mckane_PhD}, since the theory is ill-defined for all real $g$. The result for the coefficient of $\bar{g}^{2K+1}$ in $\Gamma^{(N)}( q_i , \bar{g})$ when $N$ is odd is identical to Eq.~(\ref{dis_rel_N_even_result}) if $K$ is replaced by $(2K+1)$. Of course, this is of central importance, since then $(-1)^{(2K+1)}$ is always equal to $-1$, and therefore the terms in the series for $N$ odd do not oscillate --- unlike the case with $N$ even --- and so the perturbation expansion is not Borel summable. Nevertheless we are able to write the result in a unified form for general $N>4$. We also note that here we carried out the integration on $\lambda$ first, and performed the dispersion relation afterwards. We could as well have carried out the dispersion relation first and the integration on $\lambda$ afterwards to obtain the same result, and that was in fact the way that the analysis was carried out in Ref.~\cite{mckane_PhD}.

The pole in Eq.~(\ref{dis_rel_N_even_result}) is removed by a $K^{\rm th}$ order coupling constant renormalisation, as was carried out for $\phi^4$ theory in the main text. The $\beta$ function for even $N$ can then be found to be given by
\begin{equation}
\beta(g_R) =  -\frac{(N-2)\epsilon}{2} g_R + \sum^\infty_{K>1} \beta^{(N)}_K g_R^K,
\label{beta_funct_even_N}
\end{equation}
where
\begin{eqnarray}
& & \beta^{(N)}_K = \left( - 1 \right)^K \frac{C_b(N)}{2\pi} \frac{(N-2)}{(N-1)!}\,\left[ A(N) \right]^{-(N-2)K/2} \times \nonumber \\
& & \left[ B(N) \right]^N \Gamma\left\{ \left[ K(N - 2) + (d_c + 1 + N) \right]/2 \right\}\,\left[ 1 + \mathcal{O}\left( K^{-1} \right) \right], \nonumber \\
\label{HOB_beta_even_N}
\end{eqnarray}
and where we have set $d=d_c$ following the procedure carried out in the main text in the case of $\phi^4$ theory. The result for odd $N$ is identical if $K$ is replaced everywhere by $(2K+1)$.

Although we have retained the subscript b on $C_{\rm b}(N)$, since it is formally a bare quantity, as discussed in the main text, the Gaussian fluctuations produce no divergence in $\phi^N$ theory for $N>4$, and so this quantity is finite as $\epsilon \to 0$. The fact that no renormalisation is required can also be seen from the power of $g$ which appears in the exponential in Eq.~(\ref{starting_point_phi_N_even}); a coupling-constant renormalisation carried out to lowest order only changes high-order corrections, and does not change the functional form displayed in Eq.~(\ref{starting_point_phi_N_even}). This is in contrast to the situation in $\phi^4$ theory, as discussed in the main text.

The constant $C_2(N)$, on which $C_{\rm b}(N)$ depends, is defined as a sum~\cite{mckane_PhD}, which can be evaluated in $\phi^4$ theory (or $\phi^3$ theory) where $d_c$ is an even integer, but is more difficult to find in the cases where $N>4$. Progress can be made in $\phi^6$ theory, as discussed in Ref.~\cite{mckane_PhD}, in the sense that it can be expressed in terms of a relatively simple integral which can then be evaluated. We give this result here for completeness, because it may itself be of interest~\cite{hager_2002}, and also to correct a numerical error in Ref.~\cite{mckane_PhD}:
\begin{equation}
C_2(6) = \frac{4}{\pi^2}\,\exp{ \left( - \frac{1}{\pi^2} \int^{\pi/2}_0 \,dx\,x \ln\sin x \right) }. 
\label{C_2(6)}
\end{equation}
The integral in Eq.~(\ref{C_2(6)}) comes from the evaluation of a sum of the form $\sum^\infty_{q=2} \frac{\zeta(2q-2)}{2q} z^{2q}$, which can be converted into an integral through the use of digamma functions~\cite{abramowitz_1965} or by use of a standard integral (see, for instance, Result 4.322.8 of Ref.~\cite{gradshteyn_1965}). It is related to the one given in Ref.~\cite{mckane_PhD} by an integration by parts. The integral, evaluated by Euler~\cite{broadhurst_2018}, is equal to $[7\zeta(3)-2\pi^2\ln 2]/16$~\cite{ayoub_1974}. Using this result in Eq.~(\ref{HOB_beta_even_N}) gives for $\phi^6$ theory
 \begin{equation}
\beta^{(6)}_K = \frac{2^{(49/8)} \mathcal{C}}{15\pi^4} \left( - \frac{16}{3\pi^4} \right)^K \Gamma\left( 2K + 5 \right) \left[ 1 + \mathcal{O}\left( K^{-1} \right) \right],
\label{HOB_beta_6}
\end{equation}
where $\mathcal{C} = \exp{ [ - 7\zeta(3)/16\pi^2 ] }$. 

It might be interesting to attempt an evaluation of $C_2(N)$ for other values of $N$, since having asymptotic behaviour for theories other than $N=3, 4$ or $6$, may be important, as there is current interest in conformal field theories in $d$-dimensions for values of $N$ other than these three~\cite{gracey_2017}.

\section{Derivation of two results previously utilised}
\label{sec:appB}
In this appendix we derive two results employed the main text: the form of the Fourier transform of the instanton for small argument, that is, large $\lambda$ (Eq.~(\ref{instanton_large_lambda})) and the correct form for $U_{\rm reg}$ (Eq.~(\ref{correct_U_reg})).

\subsection{The form of $\tilde{\phi}( q_i/\lambda )$ for large $\lambda$}
\label{sec:app_B1} 
Here we derive the expression given in Eq.~(\ref{instanton_large_lambda}) from Eq.~(\ref{phi_tilde}) and from the small argument expression for the Bessel function $K$. Care must be taken in applying the standard result (see Eq.~(9.6.9) of Ref.~\cite{abramowitz_1965})
\begin{equation}
K_\nu(z) \sim \frac{1}{2} \Gamma(\nu) \left( \frac{z}{2} \right)^{-\nu} \ \ \ \mathrm{Re}\,\nu > 0,
\label{naive_small_z}
\end{equation}
because additional terms diverge as $\epsilon \to 0$. To see this, recall first that (Eq.~(9.6.2) of Ref.~\cite{abramowitz_1965})
\begin{equation}
K_\nu(z) \equiv \frac{\pi}{2}\,\frac{I_{-\nu}(z) - I_\nu(z)}{\sin \left( \nu \pi \right)},
\label{K_nu_defn}
\end{equation}
and that (Eq.~(9.6.10) of Ref.~\cite{abramowitz_1965})
\begin{equation}
I_\nu(z) =  \left( \frac{z}{2} \right)^\nu \sum^\infty_{n=0} \frac{(z^2/4)^n}{n! \Gamma(\nu + n + 1)}.
\label{I_nu_expansion}
\end{equation}

Now in our case $\nu=1 - (\epsilon/2)$, which means that $\sin \left( \nu \pi \right)$ is of order $\epsilon$, indicating a potential problem --- many of the terms themselves have a pole in $\epsilon$, even though they multiply terms which make the $\lambda$ integral finite. To proceed more systematically we use the recursion formula for the $I_\mu(z)$ (see Sec.~3.71 of Ref.~\cite{watson_1944}):
\begin{equation}
I_{\mu - 1}(z) - I_{\mu + 1}(z) = \frac{2\mu}{z} I_{\mu}(z).
\label{recur_rel_mu}
\end{equation}
Now let $\mu = 1 - \nu$. Then
\begin{equation}
I_{-\nu}(z) = I_{2 - \nu}(z) + \frac{2(1-\nu)}{z} I_{1 - \nu}(z).
\label{recur_rel_nu}
\end{equation}
Using Eq.~(\ref{recur_rel_nu}), we may write Eq.~(\ref{K_nu_defn}) as
\begin{equation}
K_\nu(z) = \frac{\pi}{2}\,\frac{I_{2-\nu}(z) - I_\nu(z)}{\sin \left( \nu \pi \right)} + \frac{2(1-\nu)}{z} \frac{\pi}{2}\,\frac{I_{1-\nu}(z)}{\sin \left( \nu \pi \right)},
\label{K_nu_rewritten}
\end{equation}
or in terms of $\epsilon = 2(1-\nu)$:
\begin{eqnarray}
K_{1 - (\epsilon/2)}(z) &=& \frac{\pi}{2}\,\frac{I_{1 + (\epsilon/2)}(z) - I_{1 - (\epsilon/2)}(z)}{\sin \left( \epsilon \pi/2 \right)} \nonumber \\
&+& \frac{\epsilon \pi/2}{\sin \left( \epsilon \pi/2 \right)}\frac{I_{\epsilon/2}(z)}{z}.
\label{K_nu_rewrite}
\end{eqnarray}

We now use $I_\mu(z) = (z/2)^\mu[\Gamma(\mu + 1)]^{-1}[1 + \mathcal{O}(z^2)]$ from Eq.~(\ref{I_nu_expansion}), and write $z=q/\lambda$. In addition we multiply by $(q^2/\lambda^2)$ and also by $(q/\lambda)^{-1 + (\epsilon/2)}$ (i.e. by $(q/\lambda)^{1 + (\epsilon/2)}$) in order to obtain the $q/\lambda$ structure that is seen in the $\lambda$ integral. Doing all this we find
\begin{eqnarray}
& & \frac{\pi}{2}\,\left\{ \left( \frac{q}{\lambda} \right)^{2 + \epsilon}\left[ \frac{1}{2^{1 + (\epsilon/2)}\Gamma(2+(\epsilon/2))}\right] \right. \nonumber \\
& & \left. - \left( \frac{q}{\lambda} \right)^2 \left[ \frac{1}{2^{1 - (\epsilon/2)}\Gamma(2-(\epsilon/2))} \right] \right\}\left[ \sin \left( \epsilon \pi/2 \right) \right]^{-1} \nonumber \\ 
& & + \frac{\epsilon \pi/2}{\sin \left( \epsilon \pi/2 \right)}\frac{1}{2^{\epsilon/2}\Gamma(1 + (\epsilon/2))} \left( \frac{q}{\lambda} \right)^{\epsilon},
\label{behaviour_of_phi_tilde}
\end{eqnarray}
all multiplied by $[1 + \mathcal{O}(q/\lambda)^2]$. Now the expression in Eq.~(\ref{behaviour_of_phi_tilde}) is raised to the power $4$ in the evaluation of the $\lambda$ integral. It is clear that whenever at least one of the factors in the curly brackets of Eq.~(\ref{behaviour_of_phi_tilde}) appears, it is enough to give convergence at large $\lambda$. Also since the integrals are convergent even for $\epsilon = 0$, the two contributions from $I_{1 + (\epsilon/2)}$ and $I_{1 - (\epsilon/2)}$ cancel to leading order in $\epsilon$ and thus the denomentor is of order $\epsilon$, which cancels an $\epsilon$ from the factor $\sin ( \epsilon \pi/2 )$ in the denominator. Thus the whole contribution from the terms involving the curly brackets in Eq.~(\ref{behaviour_of_phi_tilde}) is of order one. So \textit{in effect} we may assume that 
\begin{eqnarray}
K_{1 - (\epsilon/2)}(z) \sim z^{-1} I_{\epsilon/2}(z) &\sim& z^{-1}\left( \frac{z}{2} \right)^{\epsilon/2} \frac{1}{\Gamma(1 + (\epsilon/2))} \nonumber \\
&=& z^{-1 + (\epsilon/2)}\left[ 1 + \mathcal{O}\left( \epsilon \right) \right],
\label{K_effective}
\end{eqnarray}
when wishing to find the pole in $\epsilon$ which is due to the large $\lambda$ behaviour of the $\lambda$ integrand.

\subsection{The evaluation of the constant $U_{\rm reg}$}
\label{sec:app_B2}
This constant is defined in Eq.~(\ref{U_reg_def}). Here we outline its correct evaluation~\cite{nalimov_2017,panzer_2016}. 

The function $v(k)$ is the Fourier transform of $\bar{g}_{\rm c}\bar{\phi}^2_{\rm c}/4$, in the notation of Ref.~\cite{nalimov_2001}. Therefore the first part of the sum simply involves the integral 
\begin{equation}
\int d^{d}k\,v(k) v(-k) = \int d^{d}x \frac{\bar{g}^2_{\rm c} \bar{\phi}^4_{\rm c}}{16} = - \frac{3}{2} \bar{g}_{\rm c} = 24\pi^2,
\label{first_part}
\end{equation}
in $d=4$~\cite{nalimov_2001}. 

For the second part of the sum, which involves $\ln |k|$, the procedure outlined in Ref.~\cite{nalimov_2001} is incorrect. Instead one may evaluate $v(k)$ to find
\begin{equation} 
v(k) = - \frac{3}{\pi^2} \int d^{d}x e^{i k.x} \frac{1}{(1 + x^2)^2} = - 6 K_0(k),
\label{second_part}
\end{equation}
where $K_0$ is a modified Bessel function, and $\epsilon$ has been set equal to zero in the final result. Therefore the required integral in $d=4$ equals
\begin{equation}
72\pi^2 \int^\infty_0 dk\,k^3 \ln k\,\left[ K_0(k) \right]^2.
\label{required_int}
\end{equation}
Panzer\cite{panzer_2016} has noted this integral may be carried out using Eqs.~(5) and (7) of Ref.~\cite{bailey_2008} which together give
\begin{equation}
\int^\infty_0 dk\,k^\omega \left[ K_0(k) \right]^2 = \frac{\sqrt{\pi} \Gamma^3\left(\frac{1 + \omega}{2}\right)}{4\Gamma\left( 1 + \frac{\omega}{2}\right)}.
\label{standard_int}
\end{equation}
Using this result one finds that the expression in Eq.~(\ref{required_int}) equals
\begin{equation}
12\pi^2 \left[ \frac{1}{3} + 2\ln 2 - 2\gamma \right].
\label{ln_k_int}
\end{equation}
These results when taken together give Eq.~(\ref{correct_U_reg}) of the main text.

%


\end{document}